\documentstyle[prb,aps]{revtex}
\begin{document}
\draft
\title{Freezing of anisotropic spin clusters in La$_{1.98}$Sr$_{0.02}$CuO$_4$}
\author{M. Matsuda}
\address{
The Institute of Physical and Chemical Research (RIKEN),
Wako, Saitama 351-0198, Japan}
\author{Y. S. Lee, M. Greven,$^*$ M. A. Kastner, and R. J. Birgeneau}
\address{
Department of Physics and Center for Materials Science and Engineering,\\
Massachusetts Institute of Technology, Cambridge,
Massachusetts 02139}
\author{K. Yamada$^\dagger$ and Y. Endoh}
\address{
Department of Physics, Tohoku University, Sendai 980-0812, Japan}
\author{P. B\"{o}ni}
\address{
Laboratory for Neutron Scattering, PSI, CH-5232 Villigen PSI, Switzerland}
\author{S.-H. Lee}
\address{
National Institute of Standards and Technology, NIST Center for Neutron
Research, Gaithersburg, Maryland 20899}
\author{S. Wakimoto$^\ddagger$ and G. Shirane}
\address{
Department of Physics, Brookhaven National Laboratory, Upton, New York 11973}
\date{Received 19 July 1999}
\maketitle
\begin{abstract}
A spin-glass compound, La$_{1.98}$Sr$_{0.02}$CuO$_4$, shows
quasi-three-dimensional magnetic ordering below $\sim$40 K. A remarkable feature
is that the magnetic correlation length along the orthorhombic $a\rm_{ortho}$ axis is
much longer than that along the $b\rm_{ortho}$ axis, suggesting that the spin structure
is closely related to the
diagonal stripe structure. The spin-glass state can be expressed as a random freezing
of quasi-three-dimensional spin clusters with anisotropic spin correlations
($\xi'_a\sim160$ \AA, $\xi'_b\sim25$ \AA, and $\xi'_c\sim4.7$ \AA\ at 1.6 K). The new
magnetic state is important as an intermediate phase between the three-dimensional
antiferromagnetic ordered phase in La$_2$CuO$_4$ and the incommensurate phase in
La$_{1.95}$Sr$_{0.05}$CuO$_4$ in which the positions of the incommensurate
peaks are rotated by 45$^\circ$ in reciprocal space about ($\pi$,$\pi$) from those
observed in the superconducting La$_2$CuO$_4$ compounds.
\end{abstract}
\pacs{74.72.Dn, 75.10.Jm, 75.50.Ee, 75.50.Lk}

\section{Introduction}
Extensive studies on the high-$T_c$ superconducting copper oxides have
revealed an intimate connection between the magnetism and the
superconductivity. \cite{kastner} A detailed study of the spin dynamics in the
superconducting La$_{2-x}$Sr$_x$CuO$_4$ system has been performed by Yamada
$et$ $al.$ \cite{yamada}
The phase diagram of La$_{2-x}$Sr$_x$CuO$_4$ has been explored and it has been
shown that the magnetic properties evolve dramatically with Sr doping.
The parent material La$_2$CuO$_4$ shows three-dimensional (3D) long-range
antiferromagnetic ordering below $\sim$325 K. \cite{lco,birgeneau} When Sr is
doped in the material, the 3D antiferromagnetic ordering quickly disappears and,
as originally
predicted by Aharony $et$ $al.$, \cite{aharony} the N\'{e}el state is replaced by
a spin-glass phase. In this phase elastic magnetic Bragg rods, originating from
two-dimensional (2D) spin correlations, develop gradually as shown by
Sternlieb $et$ $al.$ \cite{sternlieb} and Keimer $et$ $al.$ \cite{keimer}
In particular, Keimer $et$ $al.$ found that the magnetic peaks are almost elastic
and relatively sharp in $Q$ in La$_{1.96}$Sr$_{0.04}$CuO$_4$.
Very recently, Wakimoto $et$ $al.$ studied the magnetic properties in the spin-glass
phase (0.03$\le x\le$0.05) in detail elucidating the hole concentration dependence
of the transition temperature and the spin correlations. \cite{wakimoto}
Most importantly, they found incommensurate spin correlations in the spin-glass
La$_{1.95}$Sr$_{0.05}$CuO$_4$ in which the positions of the incommensurate
peaks were rotated by 45$^\circ$ in reciprocal space about ($\pi$,$\pi$) from those
observed in Sr-richer superconducting compounds.

On the other hand, static magnetic ordering has also been observed in superconducting
La$_{2-x}$Sr$_x$CuO$_4$. Suzuki $et$ $al.$ \cite{suzuki} and Kimura $et$
$al.$ \cite{kimura} showed that sharp incommensurate
magnetic Bragg peaks develop at low temperatures in superconducting
La$_{1.88}$Sr$_{0.12}$CuO$_4$, in close analogy with the putative stripe ordering
of holes and spins found in the La$_{2-y-x}$Nd$_y$Sr$_x$CuO$_4$
system. \cite{tranquada} Very recently, quasi-3D magnetic ordering has been observed
in superconducting La$_2$CuO$_{4+\delta}$. \cite{lee} Thus, magnetic ordering
takes place in the La$_{2-x}$Sr$_x$CuO$_4$ system with various levels of hole
doping. It is essential to clarify the nature of the static magnetic behavior
in order to understand the relationship between the spin-glass, the stripe, and the
low temperature magnetic phase in the superconducting samples.

In this paper, we report on high-resolution elastic and inelastic
neutron scattering studies in La$_{1.98}$Sr$_{0.02}$CuO$_4$. Previously,
inelastic neutron scattering measurements with medium-resolution were performed
on the same sample in order to study the magnetic excitations. \cite{matsuda} The
main purpose of the previous experiments was to compare the dynamic spin properties of
the $x$=0.02 sample with those in
La$_{1.96}$Sr$_{0.04}$CuO$_4$, in which the integrated susceptibility scales with
$E/T$. \cite{keimer} Since the time that these experiments were performed, it has
become evident that the static magnetic properties
are also important as mentioned above. Stimulated by those findings, we aimed
to clarify the static magnetic properties in the spin-glass phase of
La$_{1.98}$Sr$_{0.02}$CuO$_4$, in which the 3D antiferromagnetic long-range
ordering just disappears. The most interesting issue is how
the long-range 3D antiferromagnetic ordering disappears and the spin-glass behavior
appears with hole doping.

In this study we find that below $\sim$40 K quasi-elastic magnetic peaks develop at
the positions where magnetic Bragg peaks exist in La$_2$CuO$_4$. This means that
spin correlations develop perpendicular to the CuO$_2$ plane in addition to parallel
to it, suggesting that the 2D spin fluctuations, which exist at high temperatures,
at least in part freeze with freezing temperature enhanced due to the interplanar
interaction. Our most important findings
are that the spin correlations in the CuO$_2$ plane are anisotropic at low temperatures
and that the spin directions in the spin clusters differs from that in pure La$_2$CuO$_4$.
The spin cluster dimension along the $a\rm_{ortho}$ axis ($\sim$160 \AA) is $\sim$6 times
longer than that along the $b\rm_{ortho}$ axis ($\sim$25 \AA) at 1.6 K. From these results,
it is concluded that the spin-glass state can be described as a random freezing of
quasi-3D spin clusters with anisotropic spin correlations.

The format of this paper is as follows: The scattering geometry used in this study
is summarized in Sec. 2. Experimental details are described in Sec. 3.
The experimental results of the elastic neutron measurements are presented and the
spin structure of the short-range ordered state is summarized in Sec. 4. The
experimental results on the inelastic neutron measurements are presented in Sec. 5.
In Sec. 6 we discuss the magnetic properties of the spin-glass state.

\section{Scattering geometry}
Figure 1(a) shows the scattering geometry in the $(H,H,0)$ scattering plane in the
high-temperature tetragonal phase. La$_{1.98}$Sr$_{0.02}$CuO$_4$ undergoes
a structural phase transition from tetragonal ($I4/mmm$) to orthorhombic ($Bmab$)
structure at 485 K. Figure 1(b) shows the scattering geometry in the
$(H,K,0)$ scattering plane in the low-temperature orthorhombic phase. The structure
is slightly distorted with the $a\rm_{ortho}$ and $b\rm_{ortho}$ axes almost along the
diagonal directions of the $a\rm_{tetra}$ and $b\rm_{tetra}$ axes. Ideally, the
experiments should be performed with a single domain crystal. However, four
domains are expected to exist in real crystals since a twin structure is energetically
stable. For small rectangular samples
it is possible to prepare a single domain crystal by heating the sample above
the transition temperature and then cooling it down while applying pressure along
the $a\rm_{ortho}$ or $b\rm_{ortho}$ axis. However, the heat treatment is not efficient
for large crystals so that preparation of a single domain sample is extremely difficult
in practice. Consequently in this experiment, we were forced to use a four-domain crystal.

The scattering geometry for the four-domain crystal is shown in Fig. 1(c).
$(H,K,L)\rm_{ortho}$
and $(K,H,L)\rm_{ortho}$ nuclear Bragg peaks are observed at nearby positions. As a
result, four peaks are observed around (1,1,0)$\rm_{tetra}$ and three peaks are
observed around (2,0,0)$\rm_{tetra}$ in the $(H,K,0)$ scattering plane. Figure 2(a)
shows an elastic scan (scan A as indicated in Fig. 1(c)) at (2,0,0)$\rm_{tetra}$.
The two side peaks at
(2,0,0)$\rm_{tetra}$ originate from two of the domains and the central peak, which is
factor of $\sim$2 larger than each side peak, originates from the other two domains.
From these results, we estimate that the four domains are equally distributed.
Due to the twin structure, the $(H,0,L)\rm_{ortho}$ and the $(0,K,L)\rm_{ortho}$ scattering
planes are superposed upon each other as shown in Fig. 1(d). Scan B in Fig. 1(d)
corresponds to scan C in Fig. 1(c). Since the vertical resolution is quite broad,
the (2,0,0)$\rm_{ortho}$ ((0,2,0)$\rm_{ortho}$) peak shown in Fig. 2(b) actually originates
from two separate (2,0,0)$\rm_{ortho}$ ((0,2,0)$\rm_{ortho}$) peaks from two domains above
and below the scattering plane.

There are two ways to express Miller indices; the high temperature tetragonal phase
notation $(H,K,L)\rm_{tetra}$ and the low temperature orthorhombic phase notation
$(H,K,L)\rm_{ortho}$. Since all of the results shown in this paper are observed in the
orthorhombic phase and also obtained in the $(H,0,L)\rm_{ortho}$ and
$(0,K,L)\rm_{ortho}$ scattering planes, $(H,K,L)\rm_{ortho}$ will be used to express
Miller indices.

\section{Experimental Details}
The single crystal of La$_{1.98}$Sr$_{0.02}$CuO$_4$ was grown from a
non-stoichiometric CuO-rich solution. \cite{hidaka} The dimensions of the
plate-like shaped crystal are about 20 $\times$ 20 $\times$
3 mm$^3$. The effective mosaic of the single crystal is less than
0.5$^\circ$ full-width-at-half-maximum. The Sr concentration was determined
directly from an electron probe microanalysis measurement and indirectly from neutron
scattering measurements of the structural phase transition temperature,
$T\rm_{st}$=485 K.
The lattice constants are $a\rm_{ortho}$=5.333 \AA, $b\rm_{ortho}$=5.414 \AA,
and $c$=13.098 \AA\ ($b/a$=1.015) at 1.6 K.
The La$_{1.98}$Sr$_{0.02}$CuO$_4$ crystal is the same one as used in
Ref. \onlinecite{matsuda}. We observe that weak but sharp magnetic Bragg peaks
gradually develop below $\sim$165 K at the positions where magnetic Bragg peaks
exist in La$_2$CuO$_4$. However, the volume fraction is estimated to be less than
10\% if 0.3$\mu_B$ is assumed for the Cu moment as in an oxygen-rich
La$_2$CuO$_4$ ($T_N$=185 K). We assume that the 3D magnetic peaks come from a small
fraction of the volume where the Sr and/or oxygen concentration is below the
critical value. Since the magnetic signals which are discussed in this paper originate
from diffuse scattering, the sharp peaks from the 3D N\'{e}el phase can easily be
separated from the 2D spin glass signals by avoiding the regions around
$(1,0,even)\rm_{ortho}$ and $(0,1,odd)\rm_{ortho}$.

The neutron scattering experiments were carried out on the three-axis spectrometers
H7 and H8 at the Brookhaven High-Flux Beam Reactor, on the three-axis
spectrometer TASP at the cold neutron guide at the PSI-SINQ Facility, and
on the three-axis spectrometer SPINS at the cold neutron guide at the NIST
Center for Neutron Research.
For most of the elastic measurements, the horizontal collimator sequences were
40$'$-40$'$-S-40$'$-80$'$ and 32$'$-40$'$-S-40$'$-220$'$ with a fixed incident
neutron energy of $E_i$=5 meV.
For the inelastic measurements, the horizontal collimator sequences were
20$'$-20$'$-S-20$'$-80$'$ with $E_i$=14.7 meV and 72$'$-80$'$-S-80$'$-80$'$
with $E_f$=8 meV for lower energies ($E\le$4 meV) and
40$'$-40$'$-S-40$'$-80$'$ with $E_f$=14.7 meV for $E$=10 meV. Pyrolytic
graphite (002) was used as both monochromator and analyzer. Contamination
from higher-order neutrons was effectively eliminated using Be filters for
$E_i$=5 meV and pyrolytic graphite filters for $E_i$=14.7 meV and $E_f$=14.7 meV.
No filter was used for measurements with $E_f$=8 meV at TASP since the population
of the higher-order beam is considerably reduced.
The single crystal was oriented in the $(H,0,L)\rm_{ortho}$
and $(0,K,L)\rm_{ortho}$ scattering planes. For the elastic measurements the sample
was mounted in a helium pumped cryostat while for the inelastic measurements the
sample was mounted in a closed cycle refrigerator.

\section{Static Magnetic Properties}
\subsection{Neutron Elastic Experiments}
Figure 3(a) shows an elastic neutron scan at $(H,0,-0.3)$ at 1.6 K. A sharp
peak is observed at $H$=1. If the magnetic correlations were isotropic in the $ab$
plane and the correlations were purely two dimensional, it would be expected that
two sharp equi-intense peaks would be observed at $H$=0.985 ($K$=1) and 1.
\cite{14}
The data in Fig. 3(a) indicate that there are spin correlations along the $c$ axis
and/or that spin correlations are anisotropic. The
temperature dependence of the intensity at (1,0,$-$0.3) is shown in Fig. 3(b). The
filled circles represent data measured with $E_i$=5 meV
($\Delta E$=0.25 meV); the intensity gradually develops below $\sim$40 K. The
open circles represent the data measured with $E_i$=14.7 meV ($\Delta E$=0.9
meV); the intensity starts to increase at $\sim$80 K and the development of the
intensity is much broader. These results are consistent with those observed by
Keimer $et$ $al.$ in La$_{1.96}$Sr$_{0.04}$CuO$_4$. \cite{keimer} Specifically,
this means that the magnetic signal is quasi-elastic rather than truly elastic so that
the temperature dependence of the intensity depends on the energy window.

\subsection{Spin Structure}
Figures 4(a) and 4(b) show the $L$ dependence of the magnetic elastic peaks at
(1,0,$L$) and (0,1,$L$) at 1.6 K, respectively.
The background estimated from the high temperature data (60 K) was subtracted
so that the remaining signal is purely magnetic.
Broad peaks are observed at $(1,0,even)$ where magnetic Bragg
peaks exist in La$_2$CuO$_4$. There are some characteristic features. Firstly
the peaks are broad, indicating that the spin correlations along the $c$ axis are
short-ranged. Secondly, the magnetic intensity at $(1,0,even)$ initially increases
with increasing $L$, in contrast to the behavior found for the magnetic
Bragg intensities in pure La$_2$CuO$_4$. Lastly, the magnetic intensities at
$(1,0,L)$ are much larger than those at $(0,1,L)$. From these results, one can deduce
that spin clusters are formed in La$_{1.98}$Sr$_{0.02}$CuO$_4$.
However, the spin clusters have a different geometrical structure from that in pure
La$_2$CuO$_4$.

The magnetic Bragg intensity is proportional to

\begin{eqnarray}
\left[\mbox{\boldmath $\rm{S}$}_{\perp}f(Q)\sum_{j}
{\rm exp}(i\mbox{\boldmath $\rm{Q}$}\cdot\mbox{\boldmath $\rm{R}$}_{j})\right]^2
\label{cs}
\end{eqnarray}
where $\mbox{\boldmath $\rm{S}$}_{\perp}$, $f(Q)$, and
$\mbox{\boldmath $\rm{R}$}_{j}$
are the copper spin component perpendicular to $\mbox{\boldmath $\rm{Q}$}$,
the magnetic form factor, and the copper ion positions, respectively. In pure
La$_2$CuO$_4$, the magnetic intensities at $(1,0,even)$ are just proportional to
$f(Q)^2$, which is approximately constant for the range of $L$'s
considered here \cite{freltoft}, since the copper spins point along the $b$
axis perpendicular to the $(H0L)$ scatteing plane and
$\sum{\rm exp}(i\mbox{\boldmath $\rm{Q}$}\cdot\mbox{\boldmath $\rm{R}$}_{j})$
at $(1,0,even)$ is calculated to be constant from the 3D spin structure
with the antiferromagnetic propagation vector
$\mbox{\boldmath $\hat{\tau}$}\parallel$
$\mbox{\boldmath $\hat{a}$}\rm_{ortho}$.

The simplest model to explain the increase of the intensity with increasing $L$ along
both (1,0,$L$) and (0,1,$L$) is that the cluster antiferromagnetic spin is randomly
directed within the $ab$ plane. In this case, the intensity would vary like

\begin{eqnarray}
\left[\mbox{\boldmath $\rm{S}$}_{\perp}f(Q)\right]^2=
\frac{1}{2}\left[1+{\rm sin}^2\theta(L)\right]S^2f(Q)^2
\label{int}
\end{eqnarray}
where $\theta(L)$ is the angle that the $Q$-vector of the (1,0,$L$) or (0,1,$L$) reflection
makes with the $ab$ plane.
We will justify this model after first discussing the spatial geometry of the frozen
clusters. We should note that a result equivalent to Eq. (2) is obtained by fixing
the spin direction along $(H,H,0)$ or by assuming equal admixtures of 3D correlated
phases where the spin vector $\mbox{\boldmath $\rm{S}$}$ is along or perpendicular
to $\mbox{\boldmath $\hat{\tau}$}$($\parallel$
$\mbox{\boldmath $\hat{a}$}\rm_{ortho}$).

\subsection{Anisotropic Spin Correlations}
Figures 5(a) and 5(b) show elastic neutron scans at $(H,0,2.2)$ and $(H,0,3.2)$ at
1.6 K, respectively. The background estimated at $T$=60 K was subtracted so the remaining
scattering is entirely magnetic. A
sharp and intense peak is observed at (1,0,2.2), whereas, a broad peak is observed
at (0,1,3.2) and a sharp peak with reduced integrated intensity is observed at (1,0,3.2).
These results strongly
indicate that the spin correlations are anisotropic in the $ab$ plane, that is, the
spin correlations are longer along the $a\rm_{ortho}$ axis than along the $b\rm_{ortho}$
axis. If the magnetic correlations were isotropic in the $ab$ plane, it would be expected
that two sharp peaks would be observed at $H$=0.985 ($K$=1) and 1.
The fact that the measured intensities at $(1,0,even)$ are larger than those at
$(0,1,odd)$ is a resolution effect arising from both the broadening along $b\rm_{ortho}$
and the fact that the
coarse vertical resolution effectively integrates the peaks at $(1,0,even)$ which are
elongated perpendicular to the scattering plane as shown in the inset of Fig. 4(b).

The solid lines in Figs. 4(a), 4(b), 5(a), and 5(b) are the calculated profiles using
as the intrinsic line shape 3D squared Lorentzians convoluted with the
instrumental resolution function:
\begin{eqnarray}
\lefteqn{L(H,K,L,E)=\sum_{even,\ odd}}\nonumber\\
& &\left[\left(\frac{1}{\xi{'}_a^2(H-1)^2+\xi{'}_b^2K^2+\xi{'}_c^2(L-even)^2
+1}\right)^2
+\left(\frac{1}{\xi{'}_a^2H^2+\xi{'}_b^2(K-1)^2+\xi{'}_c^2(L-odd)^2
+1}\right)^2\right]\nonumber\\
& &\times\frac{1}{E^2+\Gamma^2}
\label{loren}
\end{eqnarray}
where $\xi'_a$, $\xi'_b$, $\xi'_c$, and $\Gamma$ represent the elastic spin
correlation lengths or cluster sizes along the $a\rm_{ortho}$ axis, $b\rm_{ortho}$ axis,
and $c$ axis and the energy width, respectively.
In order to describe spin correlations which have finite lengths
three-dimensionally, one might instead have used 3D Lorentzians:
\begin{eqnarray}
\lefteqn{L(H,K,L,E)=\sum_{even,\ odd}}\nonumber\\
& &\left(\frac{1}{\xi{'}_a^2(H-1)^2+\xi{'}_b^2K^2+\xi{'}_c^2(L-even)^2+1}
+\frac{1}{\xi{'}_a^2H^2+\xi{'}_b^2(K-1)^2+\xi{'}_c^2(L-odd)^2+1}\right)\nonumber\\
& &\times\frac{1}{E^2+\Gamma^2}.
\label{4dloren}
\end{eqnarray}
However, this function has long tails in $Q$ space and does not decay as rapidly as
the observed data do. Equation (3) describes the observed data well so that
$\xi'_a$, $\xi'_b$, and $\xi'_c$ can be estimated quite reliably. We note that the
Lorentzian squared form is the expected profile for frozen 3D random clusters with sharp
boundaries.

In the calculation the spin structure models described in Sec. 4-B,
all of which give the same $L$-dependence of the intensity, are
assumed. The parameters used are $\xi'_a$=160 \AA, $\xi'_b$=25 \AA, and
$\xi'_c$=4.7 \AA. $\Gamma$ is fixed at 0.01 meV which is determined
experimentally.
The calculation describes the observed data at $(1,0,L)$,
$(0,1,L)$, $(H,0,2.2)$, and $(H,0,3.2)$ reasonably well.
The small peaks at (1,0,$odd$) and (0,1,$even$) in Figs. 4(a) and 4(b) originate
from the tails of the broad (0,1,$odd$) and (1,0,$even$) peaks, respectively.
Surprisingly, it is found that $\xi'_a$ is about 6 times longer than $\xi'_b$.
The average distance between doped holes in the CuO$_2$ plane is $\sim$30 \AA\
in La$_{1.98}$Sr$_{0.02}$CuO$_4$, which is similar to $\xi'_b$ but much smaller
than $\xi'_a$.
$\xi'_c$=4.7 \AA\ indicates that the cluster size perpendicular to the CuO$_2$
plane is similar to the distance between nearest-neighbor CuO$_2$ planes.
A schematic figure of the spin configuration in the $ab$ plane at 1.6 K is
shown in Fig. 6. As mentioned in Sec. 4-B the spin easy axis cannot be determined
uniquely from this study, but the data are consistent with a model in which the cluster
antiferromagnetic spin direction is random in the $ab$ plane.

\section{Magnetic Excitations}
Figure 7 shows neutron inelastic scans at the energies 2, 4, and 10 meV measured
at 35 K. The spectra at 2 and 4 meV show a peak at $H$=1 and a broad tail at the
left side ($H<$1). On the other hand, the spectrum at 10 meV shows a broad and symmetric
peak centered at $H\sim 0.99$. From the results shown in Sec. 4-C,
the peak profiles at 2 and 4 meV appear to reflect the static anisotropic spin
correlations. Therefore, the spectra can be described with one sharp peak at $H$=1
and one broad peak at $H$=0.985 ($K$=1). The symmetric peak profile at 10 meV, however,
indicates that the dynamical spin correlations become isotropic at high energies.

Figure 8 shows the temperature dependence of the inelastic neutron spectra at
1.5 meV. The spectrum at 35 K is consistent with that observed at 2 meV and 35 K.
The spectrum becomes more symmetric with increasing temperature. These results
indicate that the dynamic spin correlations in the $ab$ plane are anisotropic
at both low energies and low temperatures and become more isotropic at both high
energies and high temperatures.

\section{Discussion}
The spin correlations in the spin-glass phase of 
La$_{1.98}$Sr$_{0.02}$CuO$_4$
have been clarified in this study. The
static spin correlations are not simply finite-size versions of the N\'{e}el state
spin structure in pure La$_2$CuO$_4$. The characteristic features are as follows:
(i) The spin
cluster dimensions in the $ab$ plane are highly anisotropic at low temperatures.
(ii) The anisotropic behavior has both temperature and energy dependences.
(iii) The cluster antiferromagnetic spin direction appears to be randomly oriented
within the $ab$ plane.

The anisotropic spin correlations in the $ab$ plane are suggestive of the stripe phase
found in La$_{2-y-x}$Nd$_y$Sr$_x$CuO$_4$. \cite{tranquada} However, the
spin correlations in La$_{1.98}$Sr$_{0.02}$CuO$_4$ are different from those in
La$_{2-y-x}$Nd$_y$Sr$_x$CuO$_4$. In La$_{2-y-x}$Nd$_y$Sr$_x$CuO$_4$
the hole/spin stripes run along the $b\rm_{tetra}$ ($a\rm_{tetra}$) axis
so that the magnetic domains should be elongated along the $a\rm_{tetra}$ ($b\rm_{tetra}$)
axis. These axes are rotated by 45$^\circ$ in the $ab$ plane from the $a\rm_{ortho}$ axis
along which spin correlations are longer in La$_{1.98}$Sr$_{0.02}$CuO$_4$.
Interestingly, the spin correlations in La$_{1.98}$Sr$_{0.02}$CuO$_4$ have the same
geometry as those predicted theoretically for the low concentration hole-doped system.
A Hubbard model calculation on a two-dimensional square lattice has been
performed by Schulz \cite{schulz} and by Kato $et$ $al.$ \cite{kato} They find
that a diagonally modulated spin density wave state (diagonal stripe state) is stable
when the electron density is close to half-filling. In the diagonal state, the
hole/spin stripes run along the $a\rm_{ortho}$ or $b\rm_{ortho}$ axis, which is rotated by
45$^\circ$ in the
$ab$ plane from the $a\rm_{tetra}$ ($b\rm_{tetra}$) axis along which the stripes run in
La$_{2-y-x}$Nd$_y$Sr$_x$CuO$_4$. Since the spin ordering observed in
La$_{1.98}$Sr$_{0.02}$CuO$_4$ is short-ranged, the long-range diagonal stripe
structure is not realized in this compound. However,
short-range spin ordering with anisotropic spin correlations elongated along the
$a\rm_{ortho}$ axis can be considered as a precursor phenomenon for the diagonal stripe
ordering. It also seems apparent that the anisotropic spin correlations in
La$_{1.98}$Sr$_{0.02}$CuO$_4$ are directly related to the incommensurate spin
correlations observed in La$_{1.95}$Sr$_{0.05}$CuO$_4$, in which the
incommensurate peaks are rotated by 45$^\circ$ in reciprocal space about
($\pi$,$\pi$) from those observed in Sr-rich superconducting
compounds with $x\ge$0.06. \cite{wakimoto} The
diagonal stripe state is evidently more stable in this compound.

The connection with the results in La$_{1.95}$Sr$_{0.05}$CuO$_4$ can be made quantitative.
In the diagonal stripe phase, Wakimoto $et$ $al.$ observe peaks displaced along
$b\rm_{ortho}$ by a distance $\pm\delta$ where $\delta\simeq 2\pi x/b\rm_{tetra}$ with
$x$ the Sr$^{2+}$ concentration. Very recently, Matsuda $et$ $al.$ have
found that $\delta\simeq 2\pi x/b\rm_{ortho}$ in the highly insulating
spin-glass sample La$_{1.976}$Sr$_{0.024}$CuO$_4$. \cite{matsudanew}
Therefore, for $x$=0.02, $\delta\simeq0.023$ \AA$^{-1}$. We tried to
fit the observed data in Figs. 4 and 5 with the diagonal stripe
model with $\delta$ fixed at 0.023 \AA$^{-1}$. The data can be reasonably
fitted with $\xi'_a=160$ \AA, $\xi'_b=50$ \AA, and $\xi'_c=4.7$ \AA.
It is noted that $\xi'_b$ is still short and the spin correlations are slightly anisotropic
in the CuO$_2$ plane.
Thus the measured line shape in La$_{1.98}$Sr$_{0.02}$CuO$_4$ is consistent with the
diffraction profile expected from an array of disordered stripes locally oriented along
$a\rm_{ortho}$.

As mentioned in Sec. 4-B the $L$-dependence of the magnetic intensities is well
described by a model in which the AF spin of a given cluster is randomly oriented
in the $ab$ plane. The equivalent result is obtained for the cluster staggered spin
along $a\rm_{tetra}$ or $b\rm_{tetra}$ or randomly along both $a\rm_{ortho}$ and
$b\rm_{ortho}$.
In the N\'{e}el state of pure La$_2$CuO$_4$ the spin is along $b\rm_{ortho}$ while just
above $T_N$=325 K, that is, at 328 K when the correlation length is about 800 \AA\
(Ref. \onlinecite{birgeneau})
the spin is randomly oriented in the $ab$ plane. Because of the latter result it seems
physically plausible that in the frozen spin clusters below 40 K in
La$_{1.98}$Sr$_{0.02}$CuO$_4$ of dimensions 160 \AA\ $\times$ 25 \AA\ the spin
direction would also be random. This is consistent with the fact that the net Ising
anisotropy favoring the $b\rm_{ortho}$ axis from the Dzyaloshinsky-Moriya interaction
is only about 0.1 K in energy. We should note that in all cases we assume that the
propagation vector of the AF order is along $a\rm_{ortho}$ in order to account for the
pronounced peaks at $(1,0,L)$ for $L$ $even$ alone.

One interesting and important question is why anisotropic but short-range
correlations are achieved at low temperatures in our sample of
La$_{1.98}$Sr$_{0.02}$CuO$_4$ whereas long-range
diagonal incommensurate order occurs in La$_{1.95}$Sr$_{0.05}$CuO$_4$. Here, we
speculate that the primary difference is not the hole concentration but rather the
method of crystal growth. The $x$=0.05 crystal studied by Wakimoto $et$ $al.$ was
grown with the traveling-solvent-floating-zone technique which is crucible-free
whereas the $x$=0.02 crystal studied here was grown in a platinum crucible. It is
known that in the latter case some platinum is incorporated into the crystal, that
is, Pt$^{2+}$ replaces Cu$^{2+}$. In that case, the Pt$^{2+}$ impurities would exert
an effective random field on the incipient hole stripes there by destroying the
long-range order and causing the system to break up into finite size clusters
\cite{birgeneau2} as we indeed observe experimentally.
We need further studies using a Pt-free crystal with a low hole
concentration in order to check whether the short-range correlations
originate from Pt impurities or not.

We now discuss the results of the inelastic measurements. In the previous study on
La$_{1.98}$Sr$_{0.02}$CuO$_4$, \cite{matsuda} the scaling behavior with $E/T$
of the integrated dynamical spin susceptibility was observed in the energy range
3$\le E \le$9 meV. A clear deviation from the scaling function was observed at low
temperatures in the energy range $E\le$2 meV. Similar behavior was also observed
in the crucible-grown La$_{1.96}$Sr$_{0.04}$CuO$_4$ crystal studied by Keimer $et$ $al.$
\cite{keimer} In Ref. \onlinecite{matsuda} it was argued that the
susceptibility is suppressed at low energies due to the out-of-plane anisotropy.
From our new results, it also appears to be possible that the observed susceptibility is
suppressed at low energies and low temperatures because the broad peak at
$(0,1,L)$, which originates from the short correlation length along the $b\rm_{ortho}$ axis,
was not properly integrated in the experiments of Ref. \onlinecite{matsuda}. Further
studies are needed in order to determine the
important features determining the dynamical spin susceptibility at low energies and
low temperatures.

\section*{Acknowledgments}
We would like to thank Y. Hidaka for providing us with the single crystal of
La$_{1.98}$Sr$_{0.02}$CuO$_4$. We would also like to thank V. J. Emery, K. Hirota,
K. Machida, and J. Tranquada for stimulating discussions. This
study was supported in part by the U.S.-Japan Cooperative
Program on Neutron Scattering operated by the United States
Department of Energy and the Japanese Ministry of Education,
Science, Sports and Culture and by a Grant-in-Aid for Scientific
Research from the Japanese Ministry of Education,
Science, Sports and Culture. Work at Brookhaven National
Laboratory was carried out under Contract No.
DE-AC02-98CH10886, Division of Material Science, U.S.
Department of Energy. The research at MIT was supported by the National Science
Foundation under Grant No. DMR97-04532 and by the MRSEC Program of the National
Science Foundation under Award No. DMR98-08941.

\begin{figure}
\caption{Diagrams of the reciprocal lattice in the $(H,K,0)$ scattering zone in
the tetragonal phase (a) and the orthorhombic phase (b) for a single domain crystal.
The lower figures show diagrams of the reciprocal lattice in the $(H,K,0)$ scattering
zone (c) and in the $(H,0,L)_{ortho}$ scattering zone (d) for a crystal with four
domains.}
\label{fig1}
\end{figure}

\begin{figure}
\caption{The results of elastic neutron scans A and B as shown in Figs. 1(c) and
(d), respectively.}
\label{fig2}
\end{figure}

\begin{figure}
\caption{(a) An elastic scan at (1,0,$-$0.3) at 1.6 K. The solid line is a guide to
the eye. The broken lines show the centers of the peaks (1,0,$-$0.3) and (0,1,$-$0.3)
determined from the nuclear Bragg peaks (2,0,0) and (0,2,0).
(b) Temperature dependence of the peak intensity at (1,0,$-$0.3) measured with the
two different incident neutron energies 5 meV ($\Delta E$=0.25 meV) and
14.7 meV ($\Delta E$=0.9 meV). The two sets of the data are normalized at 1.6 K.}
\label{fig3}
\end{figure}

\begin{figure}
\caption{Elastic scans along $(1,0,L)$ (a) and along $(0,1,L)$ (b) at 1.6 K.
The background intensities measured at 60 K are subtracted.
The solid lines shows the results of calculations with $\xi'_a$=160 \AA,
$\xi'_b$=25 \AA, and $\xi'_c$=4.7 \AA. The inset shows the schematic configuration
of the broad magnetic peaks. The arrows a and b show scan trajectories in (a) and
(b), respectively.}
\label{fig4}
\end{figure}

\begin{figure}
\caption{Elastic scans along $(H,0,2.2)$ (a) and along $(H,0,3.2)$ (b) at 1.6 K.
The background intensities measured at 60 K are subtracted. The broken lines show
the centers of the peaks (1,0,$L$) and (0,1,$L$) ($L$=2.2 and 3.2) determined from
the nuclear Bragg peak positions. The solid lines show the results of calculations
with $\xi'_a$=160 \AA, $\xi'_b$=25 \AA, and $\xi'_c$=4.7 \AA. The inset shows the
schematic configuration of the broad magnetic peaks. The arrows show scan
trajectories.}
\label{fig5}
\end{figure}

\begin{figure}
\caption{A possible spin structure model in the spin-glass phase of
La$_{1.98}$Sr$_{0.02}$CuO$_4$. (a) Spin easy axes
are random between spin clusters. The arrows show the
spin easy axis of a sublattice of each antiferromagnetic cluster. (b) In-plane spin
arrangement of the spin cluster at $T$=1.6 K with a spin easy axis along the
$b\rm_{ortho}$ axis.}
\label{fig6}
\end{figure}

\begin{figure}
\caption{Inelastic scans along $(H,0,-0.6)$ at 35 K as a function of energy. The solid
lines are guides to the eye. The broken lines show the centers of the
peaks (1,0,$-$0.6) and (0,1,$-$0.6) determined from the nuclear Bragg peak positions.
The inset shows the schematic configuration of the broad
magnetic peaks. The arrow shows a scan trajectory.}
\label{fig7}
\end{figure}

\begin{figure}
\caption{Inelastic scans along $(H,0,0.6)$ at 1.5 meV as a function of temperature.
The solid lines are guides to the eye. The broken lines show the centers of the peaks
(1,0,0.6) and (0,1,0.6) determined from the nuclear Bragg peak positions.}
\label{fig8}
\end{figure}

\end{document}